\documentstyle[epsf,aps]{revtex}

\begin{document}

\title{Metastable configurations of spin models on random graphs}

\author{Johannes Berg and Mauro Sellitto}

\address{Abdus Salam International Centre for Theoretical Physics, 
34100 Trieste, Italy}

\maketitle
 
\begin{abstract}
One-flip stable configurations of an Ising-model on a random graph with 
fluctuating connectivity are examined. In order to perform the quenched average 
of the number of stable configurations we introduce a global order-parameter function 
with two arguments. The analytical results are compared with numerical simulations.   
\end{abstract}

\section{Introduction}

Spin-models on random graphs have a long history in the statistical mechanics of disordered 
systems. A random graph \cite{bollobas} consists of $N$ nodes where each node is linked 
at random to a finite number of other nodes. The resulting structure is locally 
tree-like but has loops of a length $\ln(N)$. 
Associating each node $i$ with an Ising-spin variable $s_i= \pm 1$ 
and each connection with a bond $J_{ij}$ -- giving a contribution $-J_{ij} s_i s_j$ to the 
Hamiltonian -- defines a spin-model on a random graph. Loops in the graph are the source of 
frustration of the model. 

As a model of spin glasses random-graph models are 
particularly attractive \cite{vianabray}, 
since they combine the analytic accessibility within the framework of mean-field theory 
\cite{orland,mezpar1,kantersom} with the finite connectivity of short-range, 
finite-dimensional models. Furthermore, random-graph models occur in problems of 
combinatorial optimisation and 
theoretical computer science \cite{remietal,weigthartmann}, where solving 
e.g. a satisfiability or a matching problem is equivalent to finding the 
ground-state of a spin-model on a graph. Considering statistical ensembles 
of such problems with the aim of characterizing 
typical problems corresponds to defining an ensemble of random graphs.  
Interest in random-graph models has intensified lately and led to solution schemes 
beyond replica-symmetry \cite{dominicismot,remirsb,mezpar,leoneetal}. 

For fully connected systems, where the local field at each site is the sum of many random terms, 
the central limit theorem ensures that the local fields are Gaussian distributed. 
The distribution of fields may thus be characterized -- at the level of replica 
symmetry -- by two variables, the mean and the variance of this distribution. 
The free energy is thus characterized   
by a few order-parameters, which are determined self-consistently. Finite-connectivity models 
on random graphs are mean-field models where the local fields do not consist of many 
terms and consequently are not Gaussian distributed. The free energy is characterized by a 
continuous order-parameter function -- the distribution of local fields. 
A key step simplifying the replica analysis of random graphs models has been the 
introduction of a global order-parameter function \cite{dominicismot,remirsb}. 
In the case of Hamiltonians containing 2-spin, 3-spin, and higher interaction terms, 
the sums emerging from averaging over the disorder may be disentangled 
using the order parameter function 
$c(\underline{\sigma})=1/N \sum_i \delta_{\underline{\sigma} \, \underline{s}_i}$. 
This function gives the fraction of sites, whose replicated spins   
$\underline{s}_i=\{ s^a_i \}$ ($a=1 \ldots n$) are in a given configuration $\underline{\sigma}$.  

In principle any function of the spin configuration may be written as a sum of 
2-spin, 3-spin, and higher interaction terms and used as an energy function. However 
correlations between these interactions may make the resulting 
Hamiltonian intractable. In this paper we discuss how to treat random-graph models with 
Hamiltonians with a non-trivial dependence on the local magnetic field at each site. 

The central problem in this case is to find an order-parameter function to disentangle 
the result of averaging terms of the form $\exp\{ i \sum_a s_i^a J_{ij} \hat{h}_j^a \}$ 
over the disorder $J_{ij}$ 
(instead of the simpler $\exp\{ i \sum_a s_i^a J_{ij} s_j^a \}$, the variables $\hat{h}_i^a$ 
emerge from defining the local fields $h_i^a$). Furthermore such an order-parameter function ought to admit a replica-symmetric 
ansatz or other schemes and an analytic continuation $n \to 0$. In this paper we show that 
this need is answered by a two-argument order-parameter function 
$c(\underline{ \sigma},\underline{ \tau}) = 
1/N \sum_i \delta_{\underline{ \sigma}\, \underline{s}_i } 
\mbox{e}^{\underline{ \hat{h}}_i.\underline{\tau} }$. Viewing the variables 
$\hat{h}_i^a, h_i^a$ as new phase-space variables coupled to the spins, this 
order parameter function may be viewed as a discrete Fourier-transform of the 
generalisation of the usual order-parameter function $c(\underline{\sigma})$ to 
$c(\underline{\sigma},\underline{\hat{h}})$ (the local fields may be integrated out explicitly). 
 
As a concrete example we calculate the quenched average in replica-symmetry 
of the number of metastable 
configurations in one of the simplest model on a random graph, the 
ferromagnetic 2-spin model. 
For this model some results have already been obtained, 
albeit restricted to the annealed approximation, both for graphs with 
fixed connectivity \cite{deanold} and 
fluctuating connectivity \cite{bergmehta2}. For fully-connected disordered 
systems, like the Sherrington-Kirkpatrick model, the problem of metastable states 
has been dealt with in the classic 
paper by Bray and Moore \cite{braymoore}. Metastable configurations
have each spin pointing in the direction of its local field, i.e. they 
are stable (or marginally stable) against single-spin flips. As so-called inherent structures, 
such configurations play a crucial role in structural glasses 
\cite{stillinger,birolimonasson}. 
In spin-models of dense granular matter they play the role of blocked configurations 
\cite{barratkurchan,bergmehtalett,dean}. Hamiltonians which non-trivially depend on the 
local magnetisation also occur in the context of lattice-gases, e.g. where sites with 
more than a certain number of neighbouring particles are energetically penalized 
\cite{birolimezard}. Finally, a treatment of the dynamics of spin models on random 
graphs will need an order-parameter function (generating function) analogous to the one 
introduced here, since the dynamics depends explicitly on the local field at each point.  
   
The paper is organized as follows: First 
the ferromagnetic 2-spin model on a random 
graph is introduced. After discussing the annealed approximation to the problem, 
the calculation of the quenched average of the number of  
metastable configurations is outlined in section \ref{twospin}. 
Particular emphasis is given to the 2-argument 
global order-parameter function. The replica-symmetric ansatz for the 
order-parameter function is discussed and the replica-symmetric result for 
the entropy of metastable configurations is given. These expressions are evaluated 
numerically and are compared to the results of Monte-Carlo simulations and thermodynamic 
integration of an auxiliary 
model in section \ref{results}. 
The results of generalizing the problem to 3-spin models are 
given in section \ref{threespin} 
and the generalisation to models with frustrated interactions is discussed in section 
\ref{disorder}. 

\section{Metastable configurations of the 2-spin ferromagnetic model}
\label{twospin}

In the following we consider one of the simplest spin-models on a random graph, 
namely the two-spin ferromagnetic model defined by the Hamiltonian
\begin{equation}
\label{hdef}
H=-\sum_{i<j} C_{ij} s_i s_j \ , 
\end{equation}
where the sites $i=1,\ldots,N$ and $s_i=\pm 1$. 
The variable $C_{ij}=1$ with $i<j$ denotes the presence 
of a bond connecting sites $i,j$ and $C_{ij}=0$ 
denotes its absence.   
Choosing $C_{ij}=1(0)$ randomly with probability 
$c/N$ ($1-c/N$) defines an ensemble of random graphs, where the number 
of bonds connected to a site is distributed with a 
Poisson distribution of finite average $c$.
 
The condition for a (marginally) metastable configuration is that at 
each site $i$ the local 
field $h_i=\sum_j C_{ij} s_j$ obeys $h_i s_i \geq 0$. This definition implies that 
a number of neutral moves remains, as spins with zero local field may be 
flipped without a cost in energy. We choose this definition, since a quench of the 
system will in general leave a number of spins with zero local field, which are 
crucial to the subsequent dynamics \cite{barratzecchina}. Nevertheless the 
marginally stable states could be excluded easily by considering only configurations 
with non-zero local magnetic field. 

\subsection{The annealed approximation}
\label{annealed}

It is instructive to see how the average number of blocked configurations is  
calculated in the annealed approximation, where the partition function is 
averaged directly over the ensemble of random graphs, before passing to the more 
complicated case of the quenched average. 

The partition function  
configurations  $Z(\beta)$ may be written as 
\begin{equation}
\label{nblocked}
Z(\beta)=\prod_{i} \left( \sum_{s_i=\pm 1} \sum_{h_i=-\infty}^{\infty} 
\delta \left(h_i;\sum_j C_{ij}s_j \right) \Theta \left( h_i s_i \right) \right) 
\exp \left\{ \beta/2 \sum_{i} h_i s_i \right\} \ ,
\end{equation}
where $\delta(x;y)=1 \ \mbox{if} \  x=y$ and $0$ otherwise, denotes a 
Kronecker-delta and $\Theta(x)$ denotes a discrete Heaviside step-function 
with $\Theta(x)=1 \ \mbox{if} \ x \geq 0$ and $0$ otherwise. 
The function $\prod_{i} \Theta(h_i s_i)$ denotes the condition for a 
metastable configuration. 
However none of the subsequent steps of the calculation affect this function and 
it may be used to encode any function of $h_i$ and $s_i$. 
The average over the ensemble of random graphs with connectivity $c$ 
may be written as 
\begin{equation}
\label{graph_av}
\langle \langle (.) \rangle \rangle = \prod_{i<j} \int dC_{ij} 
\left[ (1-c/N) \delta(C_{ij}) + c/N \delta(C_{ij}-1) \right] (.) 
\end{equation} 
We use the integral representation $\delta \left(h_i;\sum_j C_{ij}s_j \right)=
\int_0^{2 \pi} d \hat{h}_i \exp\{ -i h_i \hat{h}_i + i \sum_{j} \hat{h}_i C_{ij} s_j \}  $ 
for the Kronecker-deltas, so the average over the disorder yields 
a term of the form 
\begin{equation}
\label{dis_av1ann}
\prod_{i<j} \left[ 1 - c/N + c/N \exp \left\{ i  \hat{h}_i s_j 
	+ i \hat{h}_j s_i \right \} \right] 
=_{\lim_{N \to \infty}} \exp \left\{ -cN/2 + c/(2N) \sum_{i,j}\mbox{e}^{ i \hat{h}_i s_j 
	+ i \hat{h}_j s_i } \right\} \ .
\end{equation}
The sum over the site labels $i$ and $j$ may be disentangled by introducing 
$c_\sigma^\tau=1/N \sum_i \delta_{s_i \sigma} e^{i \hat{h}_i \tau}$ for 
$\sigma,\tau=\pm1$. Using these $4$ order parameters, the average over the disorder may be 
written as 
\begin{equation}
\label{dis_av2ann}
\exp \left\{ -cN/2 + c/(2N) \sum_{i,j}\mbox{e}^{ i \hat{h}_i s_j 
	+ i \hat{h}_j s_i } \right\}=
\exp \left\{ -cN/2 + cN/2 \sum_{\sigma,\tau } 
c_\sigma^\tau c_\tau^\sigma \right\} \ ,
\end{equation}
where the symmetry of the exponent in (\ref{dis_av1ann}) under interchange of the 
site labels $i$ and $j$ is reflected by the symmetry of the exponent 
of (\ref{dis_av2ann}) under exchange 
of $ \sigma$ and $\tau$. 

After standard manipulations, one now easily obtains the annealed average 
of the free energy  
\begin{eqnarray}
\label{fbann}
1/N \ln \langle \langle &&Z(\beta) \rangle \rangle = \mbox{extr}_{c^{\pm}_{\pm}}
\left[ -c/2 \left( c_+^{+2} + 2 c_-^+ c_+^- + c_-^{-2} \right) - c/2 \right.\\
&&\left.	+\ln \left[ \sum_{h=1}^{\infty} 
(e^{\beta/2} \sqrt{ c_-^-/c_+^- })^h I_h( 2c \sqrt{c_-^- c_+^-} ) 
+I_0( 2c \sqrt{c_-^- c_+^-} ) + I_0( 2c \sqrt{c_+^+ c_-^+} )
+\sum_{h=1}^{\infty} 
(e^{\beta/2} \sqrt{ c_+^+/c_-^+ })^h I_h( 2c \sqrt{c_+^+ c_-^+} ) \ ,\nonumber
\right]
\right]
\end{eqnarray} 
where $I_h(x)$ denotes the modified Bessel-function of the first kind of 
order $h$. The extremum is over the order parameters $c_\sigma^\tau$. 

\subsection{The quenched average}

Using the replica 
trick $\ln Z= \lim_{n \to 0} \partial_n Z^n$ to represent the logarithm 
of the partition function, the quenched average over the ensemble of random graphs 
of the free energy of metastable configurations may be written as 
\begin{eqnarray}
\label{partition1}
\langle \langle &&Z^n(\beta) \rangle \rangle = \prod_{i<j} \int dC_{ij} 
\left[ (1-c/N) \delta(C_{ij}) + c/N \delta(C_{ij}-1) \right] \nonumber \\
&&\prod_{i,a} \left( \sum_{s_i^a=\pm 1} \sum_{h_i^a=-\infty}^{\infty} 
\delta \left(h_i^a;\sum_j C_{ij}s_j^a \right) \Theta \left( h_i^a s_i^a \right) \right) 
\exp \left\{ \beta/2 \sum_{i,a} h_i^a s_i^a \right\} \ ,
\end{eqnarray}
where the sum over the site indices $i$ is from $1$ to $N$ and the sum over the 
replica indices goes from $1$ to $n$, where $n$ is taken to be an integer. 
The Kronecker-deltas 
defining $h_i^a$ are again represented using auxiliary integrals over the auxiliary 
variables $\hat{h}_i^a$ giving 
\begin{eqnarray}
\label{partition2}
\langle \langle &&Z^n(\beta) \rangle \rangle = \prod_{i<j} \int dC_{ij} 
\left[ (1-c/N) \delta(C_{ij}) + c/N \delta(C_{ij}-1) \right] \nonumber \\
&&\prod_{i,a} \left( \sum_{s_i^a=\pm 1} \sum_{h_i^a=-\infty}^{\infty} 
\int_0^{2 \pi}d \hat{h}_i^a/(2 \pi) \right)
\exp \left\{ -i \sum_{i,a} h_i^a \hat{h}_i^a + i \sum_{i,j,a} \hat{h}_i^a C_{ij} s_j^a 
+ \beta/2 \sum_{i,a} h_i^a s_i^a \right\} 
\prod_{i,a} \Theta \left( h_i^a s_i^a \right) 
\ ,
\end{eqnarray}
so that the disorder-term $C_{ij}$ occurs only a single time in the exponent. The average 
over the term containing the disorder thus yields
\begin{equation}
\label{dis_av1}
\prod_{i<j} \left[ 1 - c/N + c/N \exp \left\{ i \sum_a \hat{h}_i^a s_j^a 
	+ i \sum_a \hat{h}_j^a s_i^a \right \} \right] 
=_{\lim_{N \to \infty}} \exp \left\{ -cN/2 + c/(2N) \sum_{i,j}\mbox{e}^{ i \sum_a \hat{h}_i^a s_j^a 
	+ i \sum_a \hat{h}_j^a s_i^a } \right\} \ .
\end{equation}
The crucial problem at this point is to find an order-parameter function suitable to 
decouple the sums over the sites in the exponent. A natural choice would be 
$1/N \sum_i \delta(\underline{ \hat{h}}-\underline{ \hat{h}_i}) 
\delta_{\underline{ \sigma} \underline{ s_i} }$ where the underlined terms 
are used to denote vectors in 
replica space, i.e. $\underline{ \sigma}=\{ \sigma^a \}$. However 
$\underline{ \hat{h}}$ is a vector of 
\emph{continuous} variables, which means that there is no simple replica-symmetric 
ansatz for this order-parameter function. Instead we generalize the order parameters 
used in the annealed approximation in section \ref{annealed} and define 
$c(\underline{ \sigma},\underline{ \tau}) = 1/N \sum_i \delta_{\underline{ \sigma}\, \underline{s}_i } 
\mbox{e}^{\underline{ \hat{h}}_i.\underline{\tau} }$, where $\underline{\tau}$ is another 
binary vector in replica space. 
Using this global order-parameter function of two arguments (\ref{dis_av1}) becomes 
\begin{equation}
\label{dis_av2}
\exp \left\{ -cN/2 + cN/2 \sum_{\underline{ \sigma},\underline{ \tau} } 
c(\underline{ \sigma},\underline{ \tau}) c(\underline{ \tau},\underline{ \sigma}) \right\} \ .
\end{equation}

After some formal manipulations outlined in the appendix one obtains as the replicated 
partition-function
\begin{eqnarray}
\label{partition3}
\langle \langle &&Z^n(\beta) \rangle \rangle = \prod_{\underline{ \sigma},\underline{ \tau}} 
	\int dc(\underline{ \sigma},\underline{ \tau})
	\exp \left\{-cN/2 \sum_{\underline{ \sigma},\underline{ \tau} } 
	c(\underline{ \sigma},\underline{ \tau}) c(\underline{ \tau},\underline{ \sigma})-cN/2
 	\right\}  \\
	&&\left[ \sum_{\underline{ \sigma},\underline{ h}} \int_0^{2 \pi} d \underline{ \hat{h}}/( 2 \pi)^n
	 \exp \left\{-i \underline{ h}.\underline{\hat{h}}+1/2 \beta \underline{ h}.\underline{\sigma} + c \sum_{\underline{ \tau} } 
	c(\underline{ \tau},\underline{ \sigma}) \mbox{e}^{i \underline{ \hat{h}}.\underline{\tau} }
	\right\} 
	\prod_a \Theta(h^a \sigma^a)		
	\right]^N \nonumber
\ .
\end{eqnarray}
The integral over the order-parameter function may be performed in the thermodynamic limit 
$N \to \infty$ as a saddle-point integral. 
Using the shorthand $\mbox{e}^\Lambda$ for the term in square brackets in (\ref{partition3}) we 
obtain the self-consistent equation for the order-parameter function
\begin{equation}
\label{spe}
c(\underline{ \sigma},\underline{ \tau})=\mbox{e}^{-\Lambda} \sum_{k=0}^{\infty} \frac{c^k}{k!}
	\prod_{l=1}^k \left( \sum_{ \underline{ \rho}_l } c(\underline{ \rho}_l,\underline{ \sigma}) \right)
	\exp \left\{ 1/2 \beta \sum_a (\tau^a+\sum_l \rho^a_l) \sigma^a
	\right\}
	\prod_a \Theta( \sigma^a (\tau^a+\sum_l \rho^a_l) ) \ , 
\end{equation}
where $\rho_l^a$ is summed over the values $\pm1$. The assumption that 
$c(\underline{ \sigma},\underline{ \tau})$ is real-valued is consistent with this equation. 
   
To treat this equation in the limit $n \to 0$ we need to make an \emph{ansatz} concerning the 
form of the order-parameter function. The simplest possible choice is the replica-symmetric (RS) 
ansatz, which assumes that the order-parameter function is invariant under the permutation 
of the replica-indices \cite{MPV}. This in turn implies that $c(\underline{ \sigma},\underline{ \tau})$ 
is a function of $\sum_a \sigma^a$, $\sum_a \sigma^a \tau^a$, and $\sum_a \tau^a$ only. 
The replica-symmetric order-parameter function may thus be written as 
\begin{equation}
\label{rsop}
c(\underline{ \sigma},\underline{ \tau})= \int dx \, dy\, dz \, P(x,y,z) \frac{\exp \left\{
	\beta x \sum_a \sigma^a + \beta y \sum_a \sigma^a \tau^a + \beta z \sum_a \tau^a 
\right\}  }
{\left[2 \mbox{e}^{\beta y} \cosh\left(\beta \left(x+z\right) \right) 
	+ 2 \mbox{e}^{-\beta y} \cosh\left(\beta \left(x-z\right) \right)   \right]^n} \ ,
\end{equation}
where the denominator serves to normalize $\int dx \, dy\, dz \, P(x,y,z) 
= \sum_{ \underline{ \sigma},\underline{ \tau} } c(\underline{ \sigma},\underline{ \tau})$. 

Inserting the RS-ansatz (\ref{rsop}) into the self-consistent equation (\ref{spe}) 
and taking the limit $n \to 0$ one finally obtains the self-consistent equation for 
$P(x,y,z)$ in the form of an invariant density
\begin{eqnarray}
\label{rsspe}
P(x,y,z)= \mbox{e}^{-c} \sum_{k=0}^{\infty} \frac{c^k}{k!}&&
	\prod_{l=1}^k \int dx_l \, dy_l \, dz_l P(x_l,y_l,z_l) \,
	\delta \left(x-\frac{1}{4 \beta} \ln\left( \frac{ f_{++} f_{+-} }{ f_{-+} f_{--} }\right) \right) \nonumber \\
	&&\delta \left(y-\frac{1}{4 \beta} \ln\left( \frac{ f_{++} f_{--} }{ f_{-+} f_{+-} }\right)\right)
	\delta \left(z-\frac{1}{4 \beta} \ln\left( \frac{ f_{++} f_{-+} }{ f_{+-} f_{--} }\right)\right)
\ ,
\end{eqnarray}
where $f_{\sigma \tau}$ serves as a shorthand for 
\begin{equation}
\label{fdef}
f_{\sigma \tau} = f( \{ x_l,y_l,z_l\},\sigma,\tau )= 
	\prod_{l=1}^k \sum_{\rho_l} \exp \left\{ \beta \sum_l x_l \rho_l + 
	\beta \sum_l y_l \rho_l \sigma + \beta \sum_l z_l \sigma + \beta/2 \sigma \sum_l \rho_l 
+\beta/2 \sigma \tau \right\} \Theta \left( \sigma \left( \tau + \sum_l \rho_l \right) \right) \ . 
\end{equation}
Without the last term - which encodes the blocking condition - this expression would factorize in 
$l$. 

From the RS order-parameter function $P(x,y,z)$ at a given value of $\beta$ 
one may also derive the values of physical observables, such as the energy, 
\begin{eqnarray}
\label{rsener}
\langle \langle \, \langle &&-\frac{1}{2N} \sum_i s_i h_i \rangle \, \rangle \rangle =
\mbox{e}^{-c} \sum_{k=0}^{\infty} \frac{c^k}{k!}
\prod_{l=1}^k \int dx_l \, dy_l \, dz_l P(x_l,y_l,z_l) \\
&&\frac{
\sum_\sigma \prod_l^k \sum_{\rho_l} (-1/2 \sigma \sum_l \rho_l) 
\exp \left\{ \beta \sum_l x_l \rho_l + \beta \sum_l y_l \rho_l \sigma + 
\beta \sum_l z_l \sigma + \beta/2 \sigma \sum_l \rho_l 
\right\} \Theta \left( \sigma \sum_l \rho_l \right)
}
{
\sum_\sigma \prod_l^k \sum_{\rho_l} \ \ \ \ \ \ \ \ \ \ \ \ \ \ \ \ \ \
\exp \left\{ \beta \sum_l x_l \rho_l + \beta \sum_l y_l \rho_l \sigma + 
\beta \sum_l z_l \sigma + \beta/2 \sigma \sum_l \rho_l 
\right\} \Theta \left( \sigma \sum_l \rho_l \right)
}
\ , \nonumber
\end{eqnarray}
or the fraction of sites with a given local field $h$ 
\begin{eqnarray}
\label{rsdeltah}
\langle \langle \, \langle &&-\frac{1}{2N} \sum_i \delta_{h;h_i} \rangle \, \rangle \rangle =
\mbox{e}^{-c} \sum_{k=0}^{\infty} \frac{c^k}{k!}
\prod_{l=1}^k \int dx_l \, dy_l \, dz_l P(x_l,y_l,z_l) \\
&&\frac{
\sum_\sigma \prod_l^k \sum_{\rho_l} (\delta_{h;\sum_l \rho_l}) 
\exp \left\{ \beta \sum_l x_l \rho_l + \beta \sum_l y_l \rho_l \sigma + 
\beta \sum_l z_l \sigma + \beta/2 \sigma \sum_l \rho_l 
\right\} \Theta \left( \sigma \sum_l \rho_l \right)
}
{
\sum_\sigma \prod_l^k \sum_{\rho_l} \ \ \ \ \ \ \ \ \ \ \ \
\exp \left\{ \beta \sum_l x_l \rho_l + \beta \sum_l y_l \rho_l \sigma + 
\beta \sum_l z_l \sigma + \beta/2 \sigma \sum_l \rho_l 
\right\} \Theta \left( \sigma \sum_l \rho_l \right)
}
\ . \nonumber
\end{eqnarray}
The single pointed brackets $\langle \rangle$ serve as a shorthand for the average 
over the metastable configurations as defined by the partition-function (\ref{partition1}). 

Finally, inserting the RS-ansatz (\ref{rsop}) into the partition-function (\ref{partition3}), 
the free energy of blocked configurations averaged over the disorder may be obtained 
\begin{eqnarray}
\label{rsfree}
\frac{1}{N} \langle \langle \ln Z \rangle \rangle &=& -\beta f(\beta) =
-c/2 \prod_{l=1}^2 \int dx_l \, dy_l \, dz_l P(x_l,y_l,z_l) \\
&&\ln \left[\frac{
\mbox{e}^{\beta(y_1+y_2)} \cosh \left( \beta \left( x_1+z_1+x_2+z_2 \right) \right) + 
\mbox{e}^{-\beta(y_1+y_2)} \cosh \left( \beta \left( x_1-z_1-x_2+z_2 \right) \right)
}{
2 \prod_{l=1}^2 \left( \mbox{e}^{\beta y_l} \cosh\left(\beta \left(x_l+z_l\right) \right) 
	+  \mbox{e}^{-\beta y_l} \cosh\left(\beta \left(x_l-z_l\right) \right) \right)
}
\right] \nonumber \\
&&+\mbox{e}^{-c} \sum_{k=0}^{\infty} \frac{c^k}{k!}
\prod_{l=1}^k \int dx_l \, dy_l \, dz_l P(x_l,y_l,z_l) \nonumber \\
&&\ln \left[\frac{
\sum_\sigma \prod_l^k \sum_{\rho_l}
\exp \left\{ \beta \sum_l x_l \rho_l + \beta \sum_l y_l \rho_l \sigma + 
\beta \sum_l z_l \sigma + \beta/2 \sigma \sum_l \rho_l 
\right\} \Theta \left( \sigma \sum_l \rho_l \right)
}{
\prod_l^k \left(2 \mbox{e}^{\beta y_l} \cosh\left(\beta \left(x_l+z_l\right) \right) 
	+ 2 \mbox{e}^{-\beta y_l} \cosh\left(\beta \left(x_l-z_l\right) \right)   \right)
} \right]
\ . \nonumber
\end{eqnarray}
Details of the calculations leading to these expressions may be found in the appendix. 

\subsection{Evaluation of the RS order-parameter function}
\label{twospinnum}

Since the order-parameter function $P(x,z,y)$ depends on three variables, the solution 
of the self-consistent equation (\ref{rsspe}) and the evaluation of 
(\ref{rsener})-(\ref{rsfree}) pose a formidable challenge. However the form of (\ref{rsspe}) 
as an invariant density suggests the use of a simple population dynamics to solve the 
self-consistent equation and evaluate the entropy of metastable configurations and other 
physical quantities. Recently this population dynamics has been used 
extensively in \cite{mezpar}. For the present problem it may be adopted as follows: 
We consider a large number ${\cal N}$ of triples labeled $i=1,\ldots,{\cal N}$ 
of numbers $\{x_i,y_i,z_i\}$ with 
$P(x,y,z)=1/{\cal N} \sum_i \delta(x-x_i) \delta(y-y_i) \delta(z-z_i) $. 
The self-consistent equation (\ref{rsspe}) may then be solved numerically 
according to the following scheme:
\begin{itemize}
\item choose an integer $k$ at random according to the Poisson distribution with 
mean $c$. 
\item choose $k$ triples at random and use them to compute $f_{--},f_{-+},f_{+-},f_{++}$ 
according to (\ref{fdef}). 
\item chose another triple $i$ at random and set $x_i= \frac{1}{4 \beta} \ln\left( \frac{ f_{++} f_{+-} }{ f_{-+} f_{--} } \right)$, $y_i=\frac{1}{4 \beta} \ln\left( \frac{ f_{++} f_{--} }{ f_{-+} f_{+-} } \right)$, $z_i=\frac{1}{4 \beta} \ln\left( \frac{ f_{++} f_{-+} }{ f_{+-} f_{--} } \right)$. 
\end{itemize}
Repeating these steps a sufficiently large number of times to ensure convergence yields an 
approximation for the order-parameter function $P(x,z,y)$, whose quality depends on the 
number of triples ${\cal N}$ used. 

In the same way the expressions (\ref{rsener})-(\ref{rsfree}) may be evaluated. Choosing 
a set of $k$ triples at random (where $k$ is sampled from the appropriate distribution) 
the integrands of (\ref{rsener})-(\ref{rsfree}) may be computed. Averaging 
the results of sufficiently many such steps we obtain approximations of the multiple 
integrals over $x,y,z$ in these expressions.   

The results discussed in the following section were obtained with ${\cal N}=5000$, 
$350000$ iterations steps, and $50000$ steps to calculate (\ref{rsener})-(\ref{rsfree}).

\subsection{Numerical simulation}
\label{numerics}

To check our analytical results - evaluated numerically - 
against the results of numerical simulations, we use a method based on 
Monte-Carlo simulation of an auxiliary model and thermodynamic integration. 
Denoting the number of 
sites with $h_i s_i \geq 0$ as $N_b$ we define the auxiliary Hamiltonian ${\cal H}$
\begin{equation}
\label{h_aux}
\beta_{\mbox{\small{aux}}} {\cal H}= \beta  H + \beta_{\mbox{\small{aux}}}(N-N_b) \ ,
\end{equation}
where $H$ is the original Hamiltonian (\ref{hdef}) and $\beta$ is a Lagrange
multiplier fixing the energy $E$ of the original system. 
By construction, the states accessed in the limit $\beta_{\mbox{\small{aux}}} \to \infty$ of 
${\cal H}$ are the blocked 
configuration of the original Hamiltonian with energy $E$ fixed by the parameter $\beta$ 
(provided they exist)\cite{jorge}. 
Numerically, blocked configurations of (\ref{hdef}) of a given energy may thus 
be generated by a Monte-Carlo dynamics of the auxiliary Hamiltonian (\ref{h_aux}), 
starting at a low auxiliary temperature 
$1/\beta_{\mbox{\small{aux}}}$, and gradually increasing $\beta_{\mbox{\small{aux}}}$ until the 
ground state is found. The auxiliary temperature must be changed sufficiently 
gradually to ensure the system always remains in equilibrium.  

Having taken the limit $\beta_{\mbox{\small{aux}}} \to \infty$ the Lagrange multiplier $\beta$ 
is simply the inverse temperature of the blocked configurations. 
Thus the entropy of 
the blocked configurations may be obtained up to a constant by thermodynamic integration 
\begin{equation}
\label{thermint}
\int_{ E(\beta=\infty)}^{ E(\beta)} 
dE \beta = 
 S(\beta)-S(\beta=\infty)
 \ .
\end{equation}

The numerical results of the following section were generated by annealing (\ref{h_aux}) 
on four realisations of a random graph with $N=1024$ increasing 
$\beta_{\mbox{\small{aux}}}$ from 0 to 10 in steps 0.001 every 2000 Monte-Carlo sweeps. 
At the end of this process the energy $E$, the fraction of sites with zero local field, 
and the magnetisation were measured during the last 250 Monte-Carlo sweeps. 
Furthermore we checked the independence of the results on the annealing rate.  
In the case of the entropy, the constant of integration was obtained simply by fitting 
the resulting curve to the analytic result. 

\section{Results for the 2-spin model}
\label{results}

\begin{figure}
\epsfysize=.3 \textwidth
  \epsffile{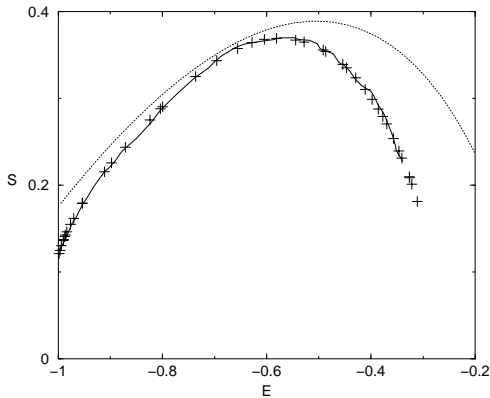}
\caption{The entropy density of metastable configurations as a function of their energy 
density $E$. The quenched result is given by the solid line, for comparison we also 
give the result of the corresponding annealed calculation from section \ref{annealed}. 
Plus-signs ($+$) denote the results of numerical simulations. At high energies 
(large negative temperatures) numerical problems with the algorithm used to solve the 
saddle-point equations arise.  
}
\label{figsc2}
\end{figure}

In the following we describe the results of the calculations of the preceding section 
for the case $c=2$ and compare them to the results of numerical simulations. 
In figure \ref{figsc2} we plot both the annealed and the quenched result for the 
entropy of blocked configurations. The results of the quenched average evaluated 
according to the algorithm of section \ref{twospinnum} fluctuate somewhat,  
so the curve is not very smooth, especially at negative temperatures.  
Nevertheless, very good agreement between the results of numerical 
simulation according to section \ref{numerics} and the quenched average of the 
entropy is found. The maximum of the entropy is reached at energies around $-.57$, so 
a randomly chosen blocked configuration will have this energy with a probability 
approaching $1$ in the thermodynamic limit. 
In the ground state $E=-1$, all 
spins of a cluster of connected points are aligned. Hence the entropy density of blocked 
configurations is simply the number of disconnected clusters of the graph times $\log(2)$. 
Using the standard results of random graph theory \cite{bollobas} we obtain a ground 
state entropy of $0.1122226...$, which agrees with the present result to within 
numerical precision.   

\begin{figure}
\epsfysize=.3 \textwidth
  \epsffile{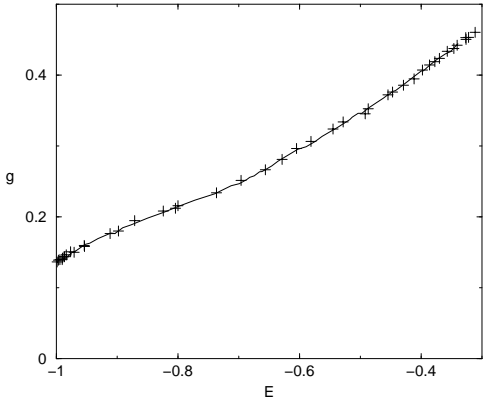}
\caption{The fraction of sites $g$ with zero local magnetic field as a function 
of the energy density $E$. Plus-signs ($+$) denote the results of numerical simulations.}
\label{figgc2}
\end{figure}

In figure \ref{figgc2} we show the fraction of sites with zero local magnetic 
field as a function of the energy density $E$, which increases monotonously with $E$. 
This effect arises since blocked sites with non-zero magnetic field give a negative 
contribution to the energy. In order to obtain blocked states also at high energies, 
the system must thus make more of the local fields equal to zero. This is also the 
reason for the decrease of the entropy at increasing energies. 

\begin{figure}
\epsfysize=.3 \textwidth
  \epsffile{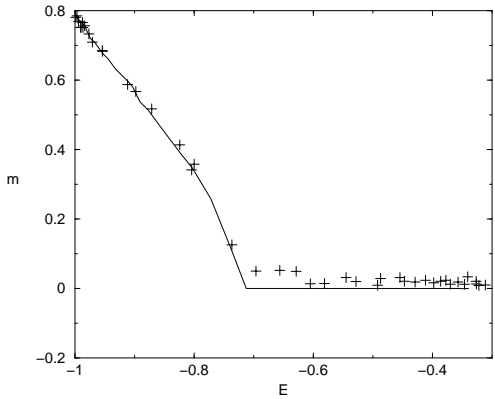}
\caption{The absolute value of the magnetisation as a function 
of the energy density $E$. Plus-signs ($+$) denote the results of numerical simulations.}
\label{figmc2}
\end{figure}

In figure \ref{figmc2} we plot the absolute value of the magnetisation of blocked 
configurations against the energy density. There is a second-order phase-transition 
from configurations with zero magnetic field at high energies, to a ferromagnetic phase 
at low energies. Again the mechanism for this is simple: in order to form blocked 
configurations at low energies, (absolutely) large local fields are required. 
These are achieved most easily by giving the system a finite magnetisation. 
This simply reflects the analogous transition in the model 
without the blocking condition, although the transition occurs at a lower energy. 
At the transition, the RS order-parameter function changes qualitatively: in the 
high-temperature phase, its weight is concentrated at $x=z=0$ and it remains a 
non-trivial function only of $y$, i.e. $P(x,y,z)=\delta(x) f(y) \delta(z)$, whereas 
in the low-temperature phase $P(x,y,z)$ is a non-trivial function of all its arguments. 

\begin{figure}
\epsfysize=.3 \textwidth
  \epsffile{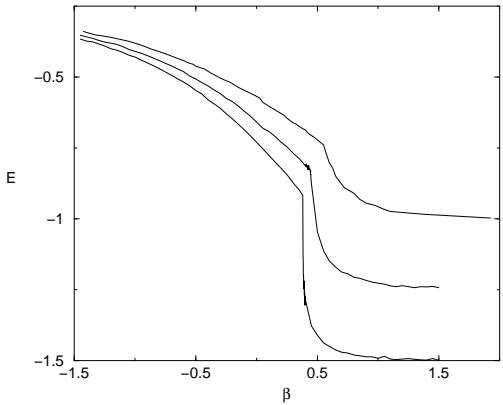}
\caption{The energy density $E$ as a function of the inverse temperature $\beta$ for 
the three connectivities $c=2,2.5,3$, from top to bottom. }
\label{figEbeta}
\end{figure}

In figure \ref{figEbeta} we plot the energy versus the inverse temperature of the 
blocked configurations $\beta$ for three connectivities $c=2,2.5,3$.   
We note that the kink in the curve -- signifying the second-order nature of the phase 
transition -- becomes more pronounced at higher connectivities. At $c=3$ it appears that 
there is a jump in the curve, which would signify that the transition had become 
first order. Furthermore, for finite running times, the algorithm solving the saddle-point 
equations shows hysteresis. However none of this is borne out by closer analysis. Near 
the transition, the three curves in figure \ref{figEbeta} collapse onto each other 
when rescaled by the width $\delta \beta$ over which the transition occurs. 
Furthermore the hysteresis in the algorithm disappears when a small magnetic field is 
applied. We thus conclude that at high connectivities, the transition remains of 
second order, but is characterized by a long plateau at the transition point.  

\begin{figure}
\epsfysize=.3 \textwidth
  \epsffile{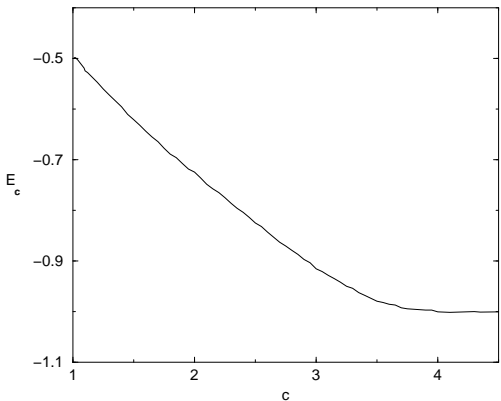}
\caption{Phase-diagram of the transition to ferromagnetic blocked configurations. 
The critical energy density $E_c$ is plotted against the average connectivity $c$.}
\label{phase_diag}
\end{figure}

Finally, in figure \ref{phase_diag} the phase diagram of the transition to 
ferromagnetic blocked states is given. 
Below the percolation threshold of the graph $c=1$ blocked states are 
typically not magnetised since the graph consists of many small disconnected clusters: 
flipping all spins of such a cluster leads from one blocked configuration to another. 
As a result $\beta_c$ diverges as $c=1$ is approached from above. Increasing $c$ 
the critical value of $\beta$ and of the corresponding energy decreases monotonously 
as expected. Nevertheless it is interesting that 
the critical energy saturates already around $c \sim 4$. 

\section{Generalisation to the three spin model}
\label{threespin}

The generalisation to Hamiltonians with 3-spin interactions is straightforward and 
we only give the results. The Hamiltonian is defined by   
\begin{equation}
\label{hdef_3}
H=-\sum_{i<j<k} C_{ijk} S_i S_j S_k 
\end{equation}
where the variable $C_{ijk}=1$ with $i<j<k$ denotes the presence 
of a plaquette connecting sites $i,j,k$ and $C_{ijk}=0$ 
denotes its absence.   
Choosing $C_{ijk}=1(0)$ randomly with probability 
$2c/N^2$ ($1-2c/N^2$) again defines an ensemble of random graphs, where 
each point is connected on average to $c$ plaquettes. 
Proceeding as in section (\ref{twospin}) we obtain for the replicated partition 
function 
\begin{eqnarray}
\label{partition3_3}
\langle \langle &&Z^n(\beta) \rangle \rangle = \prod_{\underline{ \sigma},\underline{ \tau}} 
	\int dc(\underline{ \sigma},\underline{ \tau})
	\exp \left\{-2cN/3 \sum_{\underline{ \sigma},\underline{ \tau},\underline{ \rho} } 
	c(\underline{ \sigma},\underline{ \tau \rho}) c(\underline{ \tau},\underline{ \rho \sigma}) c(\underline{ \rho},\underline{ \sigma \tau})
	-cN/3
 	\right\}  \\
	&&\left[ \sum_{\underline{ \sigma},\underline{ h}} \int_0^{2 \pi} d \underline{ \hat{h}}/( 2 \pi)^n
	 \exp \left\{-i \underline{ h}.\underline{\hat{h}}+1/3 \beta \underline{ h}.\underline{\sigma} + c \sum_{\underline{ \rho},\underline{ \eta} } 
	c(\underline{ \rho},\underline{ \sigma \eta}) c(\underline{ \eta},\underline{ \sigma \rho}) 
	\mbox{e}^{i \sum_a \hat{h}^a\rho^a \eta^a }
	\right\} 
	\prod_a \Theta(h^a \sigma^a)		
	\right]^N \nonumber
\ .
\end{eqnarray}
The self-consistent equation for $c(\underline{ \sigma},\underline{ \tau})$ is 
\begin{equation}
\label{spe_3}
c(\underline{ \sigma},\underline{ \tau})=\mbox{e}^{-\Lambda} \sum_{k=0}^{\infty} \frac{c^k}{k!}
	\prod_{l=1}^k \left( \sum_{ \underline{ \rho}_l,\underline{ \eta}_l } \right) 
	c\left(\underline{ \rho_l },\underline{ \sigma \eta_l}\right) 
	c\left(\underline{ \eta_l},\underline{ \sigma \rho_l}\right)
	\exp \left\{ 1/3 \beta \sum_a (\tau^a+\sum_l \rho^a_l \eta^a_l) \sigma^a
	\right\}
	\prod_a \Theta(\sigma^a (\tau^a+\sum_l \rho^a_l \eta^a_l) ) \ . 
\end{equation}
Using the RS ansatz (\ref{rsop}) and taking the limit $n \to \infty$ 
one obtains the self-consistent equation for $P(x,y,z)$ (\ref{rsspe}) 
\begin{eqnarray}
\label{rsspe_3}
P(x,y,z)= \mbox{e}^{-c} \sum_{k=0}^{\infty} \frac{c^k}{k!}&&
	\prod_{l=1}^k \int dx^1_l \, dy^1_l \, dz^1_l dx^2_l \, dy^2_l \, dz^2_l 
	P(x^1_l,y^1_l,z^1_l) P(x^2_l,y^2_l,z^2_l) \,
	\delta \left(x-\frac{1}{4 \beta} \ln\left( \frac{ f_{++} f_{+-} }{ f_{-+} f_{--} }\right) \right) \nonumber \\
	&&\delta \left(y-\frac{1}{4 \beta} \ln\left( \frac{ f_{++} f_{--} }{ f_{-+} f_{+-} }\right)\right)
	\delta \left(z-\frac{1}{4 \beta} \ln\left( \frac{ f_{++} f_{-+} }{ f_{+-} f_{--} }\right)\right)
\ ,
\end{eqnarray}
where
\begin{eqnarray}
\label{fdef_3}
f_{\sigma \tau} &=& f( \{ x^1_l,y^1_l,z^1_l,x^2_l,y^2_l,z^2_l\},\sigma,\tau )= 
	\prod_{l=1}^k \sum_{\rho_l,\eta_l} \exp \left\{ \beta \sum_l x^1_l \rho_l + 
	\beta \sum_l y^1_l \rho_l \eta_l \sigma + \beta \sum_l z^1_l \eta_l \sigma \right\} \\ 
	&& \exp \left\{ \beta \sum_l x^2_l \eta_l + 
	\beta \sum_l y^2_l \rho_l \eta_l \sigma + \beta \sum_l z^2_l \rho_l \sigma + 
	\beta/3 \sigma \sum_l \rho_l \eta_l 
	+\beta/3 \sigma \tau \right\} 
	\Theta \left( \sigma \left( \tau + \sum_l \rho_l \eta_l \right) \right) \ . \nonumber
\end{eqnarray}
These expressions take on the same form as their counterparts in the 2-spin case 
(\ref{spe})-(\ref{fdef}), except 
that where one term $c(\underline{ \sigma},\underline{ \tau})$ stood, there are now two --- 
a simple consequence of going from bonds connecting $2$ sites to plaquettes 
of $3$ sites. For completeness we also give the free energy of metastable configurations 
\begin{eqnarray}
\label{rsfree_3}
\frac{1}{N}&& \langle \langle \ln Z \rangle \rangle = -\beta f(\beta) =
-2c/3 \prod_{l=1}^3 \int dx_l \, dy_l \, dz_l P(x_l,y_l,z_l) \\
&&\ln \left[\frac{
\sum_{\sigma,\tau,\rho} \exp \left\{\beta \left( x_1 \sigma+y_1 \sigma \tau \rho + z_1 \tau \rho
	+ x_2 \rho + y_2 \sigma \tau \rho + z_2 \sigma \tau 
	+ x_3 \tau + y_3 \sigma \tau \rho + z_3 \rho \sigma \right) \right\}   
}{
\prod_{l=1}^3 \left(2 \mbox{e}^{\beta y_l} \cosh\left(\beta \left(x_l+z_l\right) \right) 
	+2  \mbox{e}^{-\beta y_l} \cosh\left(\beta \left(x_l-z_l\right) \right) \right)
}
\right] \nonumber \\
&&+\mbox{e}^{-c} \sum_{k=0}^{\infty} \frac{c^k}{k!}
\prod_{l=1}^k \int dx^1_l \, dy^1_l \, dz^1_l dx^2_l \, dy^2_l \, dz^2_l 
	P(x^1_l,y^1_l,z^1_l) P(x^2_l,y^2_l,z^2_l) \nonumber \\
&&\ln \left[\frac{
\sum_\sigma \prod_l^k \sum_{\rho_l,\eta_l}
\exp \left\{ \beta \sum_l\left( 
	x^1_l \rho_l +  y^1_l \rho_l \eta_l \sigma + z^1_l \eta_l \sigma 
	+ x^2_l \eta_l + y^2_l \rho_l \eta_l \sigma + z^2_l \rho_l \sigma
	+ 1/3 \sigma \rho_l \eta_l \right) \right\} 
\Theta \left( \sigma \sum_l \rho_l \eta_l \right)
}{
\prod_l^k \left(2 \mbox{e}^{\beta y^1_l} \cosh\left(\beta \left(x^1_l+z^1_l\right) \right) 
	+ 2 \mbox{e}^{-\beta y^1_l} \cosh\left(\beta \left(x^1_l-z^1_l\right) \right)   \right)
	\left(2 \mbox{e}^{\beta y^2_l} \cosh\left(\beta \left(x^2_l+z^2_l\right) \right) 
	+ 2 \mbox{e}^{-\beta y^2_l} \cosh\left(\beta \left(x^2_l-z^2_l\right) \right)   \right)
} \right]
\ . \nonumber
\end{eqnarray}
These equations may be solved in the same manner as described in section \ref{twospinnum}, 
albeit with more numerical effort. 

\section{Models with disordered bonds}
\label{disorder}

In this section we sketch how the formalism introduced in section \ref{twospin} may 
be used to cover also models with disordered bonds. 
The most prominent example of this case is the 
two-spin model on the random graph with the signs of the bonds being $\pm1$ with equal 
probability, the Viana-Bray model \cite{vianabray};  
\begin{equation}
\label{hdef_d}
H=-\sum_{i<j} J_{ij} s_i s_j \ .
\end{equation} 
The variables $J_{ij}=\pm1$ with $i<j$ denote the presence 
of a bond connecting sites $i,j$ and $J_{ij}=0$ 
denotes its absence. For $J_{ij}=\pm1$ with equal probability the average over the disorder in 
the partition function (\ref{partition1}) reads 
\begin{eqnarray}
\label{dis_av1_d}
\prod_{i<j} &&\left[ 1 - c/N + c/2N 
	\left( 
	\exp \left\{ i \sum_a \hat{h}_i^a s_j^a 
	+ i \sum_a \hat{h}_j^a s_i^a \right \}+ 
	\exp \left\{ -i \sum_a \hat{h}_i^a s_j^a 
	- i \sum_a \hat{h}_j^a s_i^a 
	\right) \right \}\right] \nonumber \\ 
&&=_{\lim_{N \to \infty}} \exp \left\{ -cN/2 + c/(4N) \sum_{i,j}
	\left(	
	\mbox{e}^{ i \sum_a \hat{h}_i^a s_j^a 
	+ i \sum_a \hat{h}_j^a s_i^a } 
	+\mbox{e}^{ -i \sum_a \hat{h}_i^a s_j^a 
	- i \sum_a \hat{h}_j^a s_i^a } 
\right) \right\} \ .
\end{eqnarray}
In order to decouple the two sums over the sites $i,j$ we may use the same order-parameter 
function $c(\underline{ \sigma},\underline{ \tau}) = 1/N \sum_i \delta_{\underline{ \sigma}\, \underline{s}_i } 
\mbox{e}^{\underline{ \hat{h}}_i.\underline{\tau} }$ defined previously and write 
for (\ref{dis_av1_d}) 
\begin{equation}
\label{dis_av2_d}
\exp \left\{ -cN/2 + c/(4N) \sum_{\underline{ \sigma},\underline{ \tau} } 
c(\underline{ \sigma},\underline{ \tau}) c(\underline{ \tau},\underline{ \sigma}) 
+
c(\underline{ \sigma},\underline{ -\tau}) c(\underline{ \tau},\underline{ -\sigma}) \right\} \ . 
\end{equation}
Eliminating the conjugate order parameter we obtain 
$i \hat{c}(\underline{ \sigma},\underline{ \tau})= c/2 \, \left(
c(\underline{ \tau},\underline{ \sigma}) +c(-\underline{ \tau},-\underline{ \sigma})\right) $. The rest 
of the calculation proceeds as in the ferro-magnetic case with the order-parameter 
function being symmetric under the simultaneous inversion of $\underline{ \sigma}$ and 
$\underline{ \tau}$

An alternative, more cumbersome route, consists in introducing a new order-parameter function   
$c(\underline{ \sigma},\underline{ \tau}) = 1/N \sum_i 
	\delta_{\underline{ \sigma} \, \underline{s}_i } 
\left( \mbox{e}^{\underline{ \hat{h}}_i . \underline{\tau} } 
+ i \mbox{e}^{-\underline{ \hat{h}}_i . \underline{\tau} } \right)$. 
Using this complex global order-parameter of two arguments (\ref{dis_av1_d}) becomes 
\begin{equation}
\label{dis_av3_d}
\exp \left\{ -cN/2 + c/(4N) \sum_{\underline{ \sigma},\underline{ \tau} } 
c(\underline{ \sigma},\underline{ \tau}) c^*(\underline{ \tau},\underline{ \sigma}) \right\} \ . 
\end{equation}
The remaining calculation proceeds exactly as in the case of the ferromagnet 
in section \ref{twospin}, the sole difference being the fact that $P(x,y,z)$ 
becomes a complex function, and equations 
(\ref{rsspe})-(\ref{rsfree}) acquire complex conjugates in the appropriate places. 
The detailed treatment however would exceed the scope of this paper. 

\section{Conclusion}

In this paper we considered the statistical mechanics of metastable configurations of 
spin models on random graphs. As a concrete example, we calculated the quenched average 
over the ensemble of random graphs of connectivity $c$ of the number of configurations with 
$h_i s_i \geq 0 \, \forall i$ for the case of the ferromagnetic $2$-spin model. 
The central tool of the calculation was a global order-parameter 
function, which unlike the standard case of Hamiltonians composed of spin-spin interactions 
of various orders, takes on the form 
$c(\underline{ \sigma},\underline{ \tau}) = 1/N \sum_i \delta_{\underline{ \sigma}\, 
\underline{s}_i } \mbox{e}^{\underline{ \hat{h}}_i.\underline{\tau} }$. The replica-symmetric 
ansatz for such an order-parameter function of two arguments was discussed and the 
saddle-point equation and the free energy were derived. 
The saddle-point equation was solved numerically using a population-dynamics algorithm. 
The results were compared in detail with numerical simulations using simulated annealing and 
thermodynamic integration. The generalisations of this 
approach to $3$-spin models and models with bond-disorder such as the Viana-Bray model 
were also discussed.

\acknowledgments
Many thanks to D. Dean, S. Franz, F. Ricci-Tersenghi, and R. Zecchina
for fruitful discussions. 

\appendix
\section{}

Here we fill in the essential steps leading to the results of section \ref{twospin}. 
The order-parameter function $c(\underline{ \sigma},\underline{ \tau})$ is introduced 
via integrals over delta-functions represented by integrals over the auxiliary variables 
$\hat{c}(\underline{ \sigma},\underline{ \tau})$. This step turns equation (\ref{partition2}) 
into 
\begin{eqnarray}
\label{partition4}
\langle \langle &&Z^n(\beta) \rangle \rangle = \prod_{\underline{ \sigma},\underline{ \tau}} 
	\int \frac{dc(\underline{ \sigma},\underline{ \tau}) d\hat{c}(\underline{ \sigma},\underline{ \tau})}{2 \pi/N} 
	\exp \left\{-iN \sum_{\underline{ \sigma},\underline{ \tau} } 
	c(\underline{ \sigma},\underline{ \tau}) \hat{c}(\underline{ \sigma},\underline{ \tau})-cN/2
	+ cN/2 \sum_{\underline{ \sigma},\underline{ \tau} } 
	c(\underline{ \sigma},\underline{ \tau}) c(\underline{ \tau},\underline{ \sigma})
 	\right\}  \\
	&&\left[ \sum_{\underline{ \sigma},\underline{ h}} \int_0^{2 \pi} d \underline{ \hat{h}}/( 2 \pi)^n
	 \exp \left\{-i \underline{ h}.\underline{\hat{h}}+1/2 \beta \underline{ h}.\underline{\sigma} + i \sum_{\underline{ \tau} } 
	\hat{c}(\underline{ \sigma},\underline{ \tau}) \mbox{e}^{i \underline{ \hat{h}}.\underline{\tau} }
	\right\} 
	\prod_a \Theta(h^a \sigma^a)		
	\right]^N \nonumber
\ .
\end{eqnarray}
The auxiliary variables may be eliminated trivially by saddle-point integration giving 
$i \hat{c}(\underline{ \sigma},\underline{ \tau})= c \, c(\underline{ \tau},\underline{ \sigma})$. Inserting this result 
into (\ref{partition4}) gives (\ref{partition3}). The self-consistent equation (\ref{spe}) 
follows directly from differentiating the exponent of (\ref{partition3}) with respect to 
$c(\underline{ \tau},\underline{ \sigma})$. The integral over $\underline{ \hat{h}}$ turns into a delta-function 
again  -- so the sum over $\underline{ h}$ may be performed -- after the exponential term 
containing the order-parameter function has been expanded as a power series. 

Now we need to insert the RS-ansatz (\ref{rsop}) into the self-consistent equation (\ref{spe}). 
Collecting all terms carrying replica-indices in the $k$th term we have
\begin{eqnarray}
\label{repterms}
\prod_l^k &&\sum_{ \underline{ \rho}_l } \exp \left\{ \beta \sum_l \left(
	x_l \sum_a \rho^a_l + y_l \sum_a \rho_l^a \sigma^a +  z_l \sum_a \sigma^a 
	+ 1/2 \sum_a \sigma^a \rho^a_l 
	\right) 
	+ \beta/2 \sum_a \sigma^a \tau^a \right\} 
	\prod_a \Theta (\sigma^a (\tau^a+\sum_l \rho^a_l) ) \nonumber \\
&&=
\prod_a \left(\prod_l^k \sum_{  \rho^a_l } 
\exp \left\{ \beta \sum_l \left( 
	x_l \rho^a_l + y_l \rho_l^a \sigma^a +  z_l \sigma^a 
	+ 1/2 \sigma^a \rho^a_l \right) 	
	+ \beta/2 \sigma^a \tau^a \right\}
\Theta (\sigma^a (\tau^a+\sum_l \rho^a_l) )
\right)  \\
&&=\exp \left\{\sum_a \ln f_{\sigma^a \tau^a} \right\} \nonumber \\
&&=\exp \left\{ 1/4 \left( \sum_a \sigma^a + \sum_a \sigma^a \tau^a + \sum_a \tau^a + n \right)
	\ln f_{1\, 1}
	+ 1/4 \left( \sum_a \sigma^a - \sum_a \sigma^a \tau^a - \sum_a \tau^a + n \right)
	\ln f_{1\, -1} \right. \nonumber \\
	&&\left. \ \ \ \ \ + 1/4 \left( -\sum_a \sigma^a - \sum_a \sigma^a \tau^a + \sum_a \tau^a + n \right)
	\ln f_{-1\, 1}
	+ 1/4 \left( -\sum_a \sigma^a + \sum_a \sigma^a \tau^a - \sum_a \tau^a + n \right)
	\ln f_{-1\, -1} \nonumber
	\right\} \ ,
\end{eqnarray}
where $f_{\sigma \tau}$ is defined in (\ref{fdef}). In the last step the we used 
\begin{equation}
\sum_a \delta_{\sigma \sigma^a}\delta_{\tau \tau^a} =\left\{ 
\begin{array}{ll}
1/4\left( \sum_a \sigma^a + \sum_a \sigma^a \tau^a + \sum_a \tau^a + n \right) & 
\mbox{for } \tau=\sigma=1 \\
 1/4\left( \sum_a \sigma^a - \sum_a \sigma^a \tau^a - \sum_a \tau^a + n \right) & 
\mbox{for } \tau=-1 \ \sigma=1 \\
 1/4\left( -\sum_a \sigma^a - \sum_a \sigma^a \tau^a + \sum_a \tau^a + n \right) & 
\mbox{for } \tau=1 \ \sigma=-1 \\ 
1/4 \left( -\sum_a \sigma^a + \sum_a \sigma^a \tau^a - \sum_a \tau^a + n \right) & 
\mbox{for } \tau=\sigma=-1 \\
\end{array} \right. \ .
\end{equation}
Writing $is=\sum_a \sigma^a$, $iu=\sum_a \sigma^a \tau^a$, and $it=\sum_a \tau^a$ the limit 
$n \to 0$ may be taken. Collecting all 
terms in $k$ and Fourier-transforming with respect to $s,u,t$ we obtain the self-consistent 
equation in RS in the form (\ref{rsspe}). The calculation of (\ref{rsener})-(\ref{rsfree}) 
follow the same scheme.

\end{document}